\documentclass[preprint2]{aastex}

\shorttitle{The {\em local} mass-metallicity relation of \hh regions}
\shortauthors{Rosales-Ortega et al.}

\usepackage{txfonts}
\usepackage{natbib}
\usepackage{amssymb}
\usepackage[latin1]{inputenc}
\usepackage{subfigure}
\usepackage{fancyhdr}
\usepackage{placeins}
\usepackage{graphicx}
\usepackage[ps2pdf,colorlinks,breaklinks=true]{hyperref}
\hypersetup{bookmarksopen,citecolor={blue},linkcolor={blue},urlcolor={blue}}

\usepackage[left=2.2cm,right=1cm,top=1.5cm,bottom=1.8cm,head=0cm,foot=0.9cm]{geometry}

\DeclareRobustCommand{\ion}[2]{%
\relax\ifmmode
\ifx\testbx\f@series
{\mathbf{#1\,\mathsc{#2}}}\else
{\mathrm{#1\,\mathsc{#2}}}\fi
\else\textup{#1\,{\mdseries\textsc{#2}}}%
\fi}

\newcommand{\hii}{\ion{H}{ii}}
\newcommand{\hh}{\ion{H}{ii}~}

\newcommand{\nii}{[\ion{N}{ii}]}

\newcommand{\oii}{[\ion{O}{ii}]}
\newcommand{\oiii}{[\ion{O}{iii}]}

\newcommand{\lam}{$\lambda$}
\newcommand{\upstar}{$^{\star}$} 
 
\newcommand{\ha}{H$\alpha$} 
\newcommand{\hb}{H$\beta$}

\newcommand{\mz}{{\small $\mathcal{M}$-Z}}
\newcommand{\mew}{{\small $\mathcal{M}$-EW(H$\alpha$)}}
\newcommand{\mze}{{\small $\mathcal{M}$-Z-EW(H$\alpha$)}}
\newcommand{\ewha}{$|$EW(\ha)$|$}

\defcitealias{RosalesOrtega:2010p3836}{Ros10}
\defcitealias{Tremonti:2004p1138}{T04}
\defcitealias{Kewley:2008p1394}{K08}

\begin{document}

\title{A new scaling relation for \hii\ regions in spiral galaxies: \\
unveiling the true nature of the mass-metallicity relation$^\star$}



\author{
  F.\,F. Rosales-Ortega\altaffilmark{1,2}, 
  S.\,F. Sánchez\altaffilmark{2,3}, 
  J. Iglesias-Páramo\altaffilmark{2,3},
  A.\,I. Díaz\altaffilmark{1},\\[4pt]
  J.\,M. Vílchez\altaffilmark{2},
  J. Bland-Hawthorn\altaffilmark{4},
  B. Husemann\altaffilmark{5},
  and D. Mast\altaffilmark{2}.
}

\altaffiltext{1}{~Departamento de F{\'i}sica Te{\'o}rica, Universidad
  Aut\'onoma de Madrid, 28049 Madrid, Spain. \texttt{frosales@cantab.net}}

\altaffiltext{2}{~Instituto de Astrof\'{\i}sica de Andaluc\'{\i}a (CSIC),
  Camino Bajo de Hu\'etor s/n, Aptdo. 3004, E18080-Granada, Spain.}

\altaffiltext{3}{~Centro Astron{\'o}mico Hispano Alem{\'a}n, Calar Alto,
  CSIC-MPG, C/Jes{\'u}s Durb{\'a}n Rem{\'o}n 2-2, E-04004 Almeria, Spain.}

\altaffiltext{4}{~Sydney Institute for Astronomy, School of Physics A28,
  University of Sydney, NSW 2006, Australia.}

\altaffiltext{5}{~Leibniz-Institut f\"ur Astrophysik Potsdam (AIP), An der
  Sternwarte 16, D-14482 Potsdam, Germany.}

\altaffiltext{\upstar}{Based on observations collected at the Centro
  Astronómico Hispano-Alemán (CAHA) at Calar Alto, operated jointly by the
  Max-Planck Institut für Astronomie and the Instituto de Astrofísica de
  Andalucía (CSIC).}

\begin{abstract}
  We demonstrate the existence of a {\em local} relation between galaxy
  surface mass density, gas metallicity, and star-formation rate density
  using spatially-resolved optical spectroscopy of \hh regions in the
  local Universe. 
  One of the projections of this distribution, --the {\em local}
  mass-metallicity relation-- extends over three orders of magnitude in galaxy
  mass density and a factor of eight in gas metallicity.
  We explain the new relation as the combined effect of the differential radial
  distributions of mass and metallicity in the discs of galaxies, and a
  selective star-formation efficiency.
  We use this local relation to reproduce --with remarkable agreement-- the
  total mass-metallicity relation seen in galaxies, and conclude that the latter is
  a scale-up integrated effect of a local relation, supporting the
  {\em inside-out} growth and {\em downsizing} scenarios of galaxy evolution.
\end{abstract}

\keywords{Galaxies: abundances --- Galaxies: fundamental parameters --- Galaxies: ISM --- Galaxies: stellar content --- Techniques: imaging spectroscopy\\}

\section{Introduction}
\label{sec:intro}

The existence of a strong correlation between stellar mass and gas-phase
metallicity in galaxies is a well known fact \citep{Lequeux:1979p4157}.
These parameters are two of the most fundamental physical properties of
galaxies, both directly related to the process of galaxy evolution.
The mass-metallicity (\mz) relation is consistent with more massive galaxies
being more metal-enriched. It was established observationally by 
\citet[][hereafter T04]{Tremonti:2004p1138}, who found a tight
correlation spanning over 3 orders of magnitude in mass and a factor of
10 in metallicity, using a large sample of star-forming galaxies up to
z\,$\sim$\,0.1 from the Sloan Digital Sky Survey (SDSS). The \mz\ relation appears
to be independent of large-scale environment \citep{Mouhcine:2007p4175} and
has been established at all accessible redshifts \citep[e.g.][]{Savaglio:2005p4162,Erb:2006p4161,Maiolino:2008p3862}.

Considerable work has been devoted to understanding the physical mechanisms
underlying the \mz\ relation. The proposed scenarios to explain its origin can
be broadly categorized as: 1) the loss of enriched gas by outflows
(T04; \citealt{Kobayashi:2007p4169}); 2) the accretion of 
pristine gas by inflows \citep{Finlator:2008p4158}; 3) variations of the initial
mass function with mass \citep{Koppen:2007p4154}; 4) selective star formation
efficiency or {\em downsizing} \citep{Brooks:2007p4155,Ellison:2008p4160,Calura:2009p4156,ValeAsari:2009p4167};
or a combination of them. Recent studies also show evidence of a relation with
star formation rate (SFR) inferring a fundamental \mz-SFR relation
\citep[hereafter FMR,][]{LaraLopez:2010p4165,Mannucci:2010p4163,Yates:2012p4168}.

As yet, there has been no major effort to test the \mz\ relation
using spatially resolved information. 
The most likely example might be ascribed to
\citet{Edmunds:1984p223} and \citet{VilaCostas:1992p322} who noticed a correlation
between surface mass density and gas metallicity in a number of galaxies.
Nowadays, the imaging spectroscopy technique can
potentially prove key to understanding many of the systematic effects that
hamper the role of the distribution of mass and metals in galaxies. 
In this work, we use integral field spectroscopy (IFS) observations of a
sample of nearby galaxies to demonstrate: 
i) the existence of a {\em local} \mz-star-formation relation; 
and ii) how the global \mz\ relation seen in galaxies can be reproduced by the
presence of the local one. We present convincing evidence that the observed \mz\
relations represent a simple sequence in astration.

\begin{figure*}[ht]
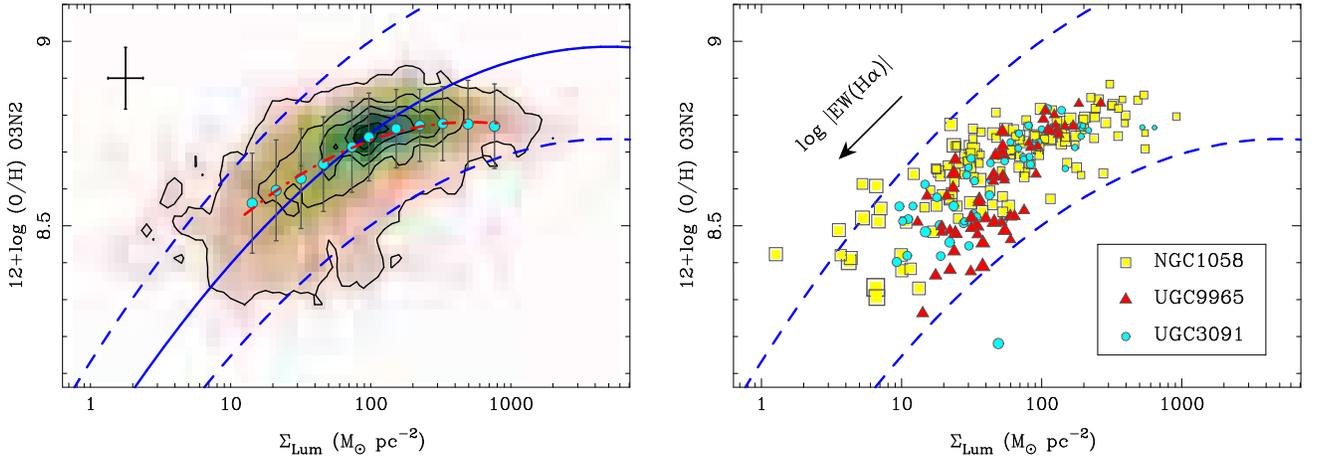

  \centering
  \includegraphics[width=6cm,angle=270]{fig1a.ps}\hspace{0.5cm}
  \includegraphics[width=6cm,angle=270]{fig1b.ps}
  \caption{
    {\em Left-panel}: The relation between surface mass density and
    gas-phase oxygen metallicity for $\sim$2000 \hh regions in nearby
    galaxies, the {\em local} \mz\ relation.
    The first contour stands for the mean density value, with a regular
    spacing of four time this value for each consecutive contour. The blue
    circles represent the mean (plus 1$\sigma$ error bars) in bins of
    0.15\,dex. The red dashed-dotted line is a polynomial fit to
    the data. The blue-lines correspond to the \citetalias{Tremonti:2004p1138}
    relation ($\pm$0.2\,dex) scaled to the relevant units. Typical errors for
    $\Sigma_{\rm Lum}$ and metallicity are represented.
    {\em Right-panel}: Distribution of \hh regions along the local \mz\
    relation for three galaxies of the sample at different redshifts. The size
    of the symbols are linked to the value of \ewha, being inversely
    proportional to $\Sigma_{\rm Lum}$ and metallicity as shown.
    \label{fig:local}
}
\end{figure*}

\section{Data sample and analysis}
\label{sec:data}

The study was performed using IFS data of a sample of
nearby disc galaxies, part belonging to the PINGS survey \citep{RosalesOrtega:2010p3836},
and a sample of face-on spiral galaxies from \citet{MarmolQueralto:2011p4103} 
as part of the feasibility studies for the CALIFA survey \citep{Sanchez:2012p4186}.
The observations were designed to obtain continuous coverage spectra of the
whole surface of the galaxies. 
The final sample comprises 38 objects, with a redshift range between
$\sim$0.001-0.025. Although this sample is by no means a statistical
subset of the galaxies in the local Universe, it is a representative sample
of face-on, mostly quiescent, spiral galaxies at the considered redshift range. 
They were observed with the PMAS spectrograph \citep{Roth:2005p2463} 
in the PPAK mode \citep{Verheijen:2004p2481,Kelz:2006p3341} on the 3.5m
telescope in Calar Alto with similar setup, resolutions and integration
times, covering their optical extension up to $\sim$2.4 effective radii within
a wavelength range $\sim$3700-7000~\AA.
Data reduction was performed using {\sc r3d} \citep{Sanchez:2006p331},
obtaining as an output a datacube for each galaxy, with a final
spatial sampling between 1-2 arcsec/pixel, which translates to a linear
physical size between a few hundreds of parsecs to $\sim$1\,kpc (depending on
the size of the object).
Details on the sample, observing strategy, setups, and data reduction can be
found in \citet{RosalesOrtega:2010p3836}, and \citet{MarmolQueralto:2011p4103}.

The \hh regions in these galaxies were detected, spatially
segregated, and spectrally extracted using {\sc HIIexplorer} \citep{Sanchez:2012b}.
We detected a total of 2573 \hh regions with good spectroscopic quality. This
is by far the largest spatially-resolved, nearby spectroscopic \hh region
survey ever accomplished.
Note that for the more distance galaxies the segregated \hh regions may
comprise a few classical ones, i.e. \hh aggregations;
which may not be useful to analyze additive/integrated
properties as in individual \hh regions (e.g. H$\alpha$ luminosity function),
but are perfectly suited for the study of line ratios and chemical abundances.
The emission lines were decoupled from the underlying stellar population
using {\sc fit3d} \citep{Sanchez:2007p3299}, following a robust and
well-tested methodology \citep{RosalesOrtega:2010p3836,Sanchez:2011p3844}.
Extinction-corrected, flux intensities of the stronger emission lines were
obtained and used to select only star-forming regions based on typical BPT
diagnostic diagrams \citep{Baldwin:1981p3310} using a combination of the
  \citet{Kewley:2001p3313} and \citet{Kauffmann:2003p3918} demarcation curves. 
Our final sample comprises 1896 high-quality, spatially-resolved \hh
regions/aggregations of disc galaxies in the local Universe.
Details on the procedure can be found in \citet{Sanchez:2012b}.

Gas-phase oxygen abundances were estimated using the O3N2 calibrator \citep{Pettini:2004p315},
based on the \oiii, \hb, \nii, and \ha\ emission lines.
We use this indicator because it is less dependent on dust attenuation and we
could use the whole \hh region sample (given the lack of \oii\ \lam3727 in
some galaxies due to redshift). 
We used the prescriptions given by \citet{Bell:2001p210} to convert $B-V$ colors into a
$B$-band mass-to-light ratio ($M/L$) to derive the (luminosity) surface mass
density ($\Sigma_{\rm Lum}$, M$_{\odot}$ pc$^{-2}$) within the area
encompassed by our IFS-segmented \hh regions. The $B$ and $V$-band surface
brightness within the considered areas were derived directly from the
(emission-line free) IFS data. 
The estimated error in the derived $\Sigma_{\rm Lum}$ is $\sim$0.25\,dex, taking
into account the error in the flux calibration of our IFS dataset
\citep{RosalesOrtega:2010p3836,MarmolQueralto:2011p4103}, and the expected
uncertainty of the \citet{Bell:2001p210} formulation.
The relation between the mass surface density and radius was checked by
normalizing the mass profile by the effective radius, finding a similar slope
for all galaxies closer to unity, i.e. the mass profile follows an exponential
law which is what is expected for disc galaxies \citep[e.g.][]{Bakos:2008p4177}.
Furthermore, we compared our radial mass profiles with the more
sophisticated $K$-band derived profiles of those common galaxies in our sample
with the DiskMass Survey (\citealt{Bershady:2010p4172}; Martinsson 2011, PhDT)
finding an agreement within 20\%, strengthening the validity of our derived
masses.

\section{The local \mz\ relation}
\label{sec:local}

\begin{figure}[t]
  \centering
  \href{http://www.ast.cam.ac.uk/ioa/research/pings/html/public/local-mz}{
    \fbox{\includegraphics[width=0.95\columnwidth,trim = 35mm 15mm 10mm 25mm, clip]{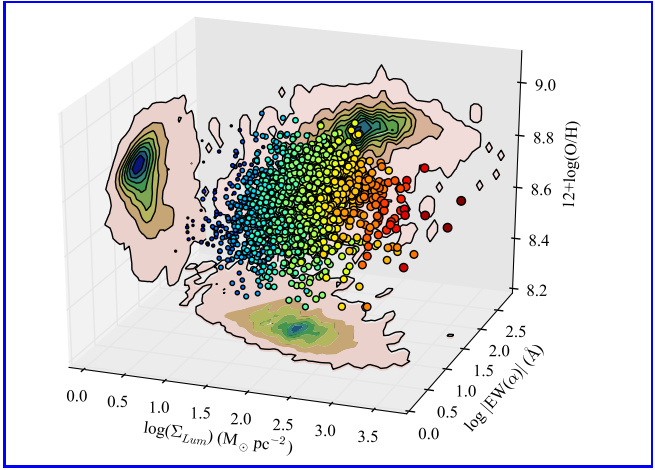}}
  }
  \caption[Caption]
  {
    3D representation of the local $\mathcal{M}$-Z-EW(H$\alpha$) relation. 
    The size and color scaling of the data points are linked to the value of
    $\log \Sigma_{\rm Lum}$ (i.e. low-blue to high-red values). 
    The projection of the data over any pair of axes reduces to the
    local \mz, \mew, and metallicity-EW(H$\alpha$) relations.
    Click on the figure for an online 3D  animated version\footnotemark.
    \label{fig:3d}
  }
\end{figure}

The left-panel of Fig. \ref{fig:local} shows the striking 
correlation between the local surface mass density and gas metallicity for our
sample of nearby \hh regions, i.e. the {\em local} \mz\ relation, extending
over $\sim$3 orders of magnitude in $\Sigma_{\rm Lum}$ and a factor $\sim$8 in metallicity.
As in the case of the global \mz\ relation \citepalias[e.g.][]{Tremonti:2004p1138},
the correlation is nearly linear for lower $\Sigma_{\rm Lum}$, flattening
gradually at higher values. 
Also remarkable is the tightness of the correlation, the 1$\sigma$ scatter
of the data about the median is $\pm$0.14 dex.
The notable similarity with the global \mz\
relation can be visually recognised with the aid of the blue-lines which stand
for the \citetalias{Tremonti:2004p1138} fit ($\pm$0.2\,dex) to the global \mz\
relation, shifted arbitrarily both in mass ($\sim$5\,mag, to account for the
difference in size between galaxies and \hh regions) and metallicity
($-$0.15\,dex, due to the well-known effect of metallicity scale) to coincide
with the peak of the \hh region \mz\ distribution, 
which clearly follows the \citetalias{Tremonti:2004p1138} relationship.
With this offset, the local \mz\ values stand within 90\% of the 95\% range of
the \citetalias{Tremonti:2004p1138} relation.
A polynomial fit of the $\Sigma_{\rm Lum}$ vs. metallicity relation yields:
\begin{eqnarray}
  12 + \log({\rm O/H})_{\rm O3N2} &=&  8.079_{\pm 0.141} +  0.525_{\pm 0.143} \Sigma_{\rm Lum} \nonumber\\
                       & & -0.098_{\pm 0.035} [\Sigma_{\rm Lum}]^2,
  \label{eq:1}
\end{eqnarray}
\noindent
valid over the range $1 < \log \Sigma_{\rm Lum} < 3$.
We obtain the same shape of the relation (and similar fit) if we adopt the
R$_{23}$ metallicity calibration of \citetalias{Tremonti:2004p1138}, but with
a higher scatter ($\sim$20\%) due to the lowest number of \hh regions in our
sample with the \oii~\lam3727 line. 
Other typical calibrations were tested finding similar results
\citep[e.g. N2][]{Denicolo:2002p361}, i.e. the scale of the relation changes
when using different indicators, but the overall, qualitative shape does not change.

\footnotetext{~Available at: \url{http://tinyurl.com/local-MZ-relation}}

Note that the local \mz\ relation holds for individual
galaxies with a large-enough dynamical range in $\Sigma_{\rm Lum}$ and
metallicity to cover the whole parameter space of the relation, i.e. it is
indeed a {\em local} relationship. This is shown in the right-panel of
Fig. \ref{fig:local}, displaying the local \mz\ relation for the \hh regions
of three galaxies in our sample at different redshifts. For the closest galaxy 
(NGC\,1058, z$\sim$0.0017), the relationship holds for $\sim$3 mag
in $\Sigma_{\rm Lum}$, and even with the loss of spatial resolution at higher
redshifts the relation persists over a large dynamic range.

\begin{figure}[t]
  \centering
  \includegraphics[width=6cm,angle=270]{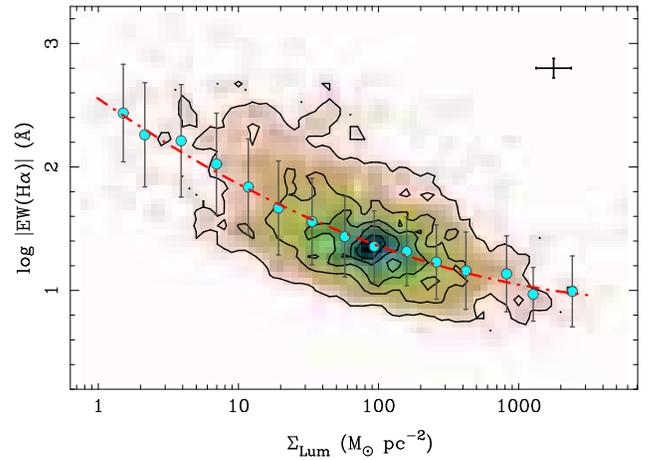}
  \caption{
    Relation between \ewha\ and surface mass density, i.e. a projection of the
    local \mze\ relation. Contours and symbols as in Fig. \ref{fig:local}. The
    red line is a polynomial fit to the data.
    \label{fig:ewha}
  }
\end{figure}

In addition, we find the existence of a more general relation between
mass surface density, metallicity, and the equivalent width of \ha, defined 
as the emission-line luminosity normalized to the adjacent continuum flux, 
i.e. a measure of the SFR per unit luminosity \citep{Kennicutt:1998p3370}.
This functional relation is evident in a 3D space with orthogonal
coordinate axes defined by these parameters, consistent with \ewha\
being inversely proportional to both $\Sigma_{\rm Lum}$ and metallicity, as
shown in Fig. \ref{fig:3d}.
The projection of the {\em local} \mze\ relation on the planes defined
in this 3D space correspond to the local \mz, \mew, and metallicity-EW(\ha)
relations. Fig. \ref{fig:ewha} shows the projection of $\log$\,\ewha\ as a strong and
tight function of $\Sigma_{\rm Lum}$ extending over more than 3 orders of
magnitude. The projection on the \ewha\ vs. metallicity plane shows an existent but
weaker correlation.

\section{The global \mz\ relation}
\label{sec:global}

In order to test whether the global \mz\ relation observed by
\citetalias{Tremonti:2004p1138} using SDSS data is a reflection (aperture
effect) of the local \hh region mass-density vs. metallicity relation, we
perform the following exercise. We simulate a galaxy with typical $M_B$ and
$B-V$ values drawn from flat distributions in magnitude ($-$15\,to\,$-$23) and
colour ($\sim$0.4$-$1). A redshift is assumed for the mock galaxy, drawn from
a Gaussian distribution with mean $\sim$0.1 and $\sigma=0.05$, 
with a redshift cut $0.02 < z < 0.3$ in order to resemble the SDSS
\citetalias{Tremonti:2004p1138} distribution.
The mass of the galaxy is derived using the integrated $B$-band
magnitudes, $B-V$ colours and the average $M/L$ ratio following \citet{Bell:2001p210}.
The effective radius of the mock galaxy is estimated using the
well-known luminosity-scale relation \citep[e.g.][]{Brooks:2011p4178}
assuming a normal standard deviation of 0.3 dex \citep{Shen:2003p4179}.
Once the mass and effective radius are known, the
surface brightness at the center of the mock galaxy is derived assuming
an exponential light distribution. Then, the metallicity is calculated at
different radii up to an aperture equal to the SDSS fiber (3 arcsec) in 100
bins using Eq.\,\ref{eq:1}, i.e. the metallicity that corresponds to the mass
density surface at each bin. The metallicity assigned to the mock galaxy is
the mean value determined within the aperture, assuming an error equal
to the standard deviation of the derived distribution plus a systematic
error of $\sim$0.1\,dex, intrinsic to the derivation of the metallicity.
We used the O3N2 metallicity conversion by \citet[][hereafter K08]{Kewley:2008p1394}
to convert to the \citetalias{Tremonti:2004p1138} base.
The process is repeated over 10,000 times in order to obtain a
reliable distribution in the mass and metallicity of the mock galaxies.

\begin{figure}[t]
  \centering
  \includegraphics[width=0.35\textwidth,angle=270]{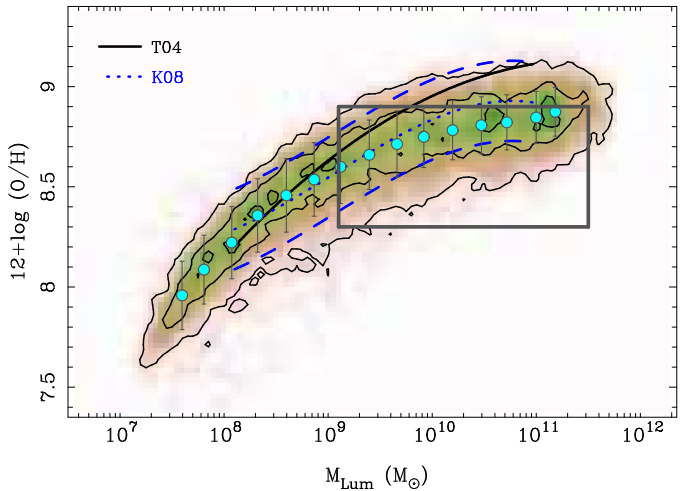}
  \caption{
    Distribution of simulated galaxies in the \mz\ plane assuming a 
    {\em local} \mz\ relation and considering the aperture effect of the SDSS
    fiber, as explained in the text.
    The contours correspond to the density of points, while the circles
    represent the mean value (plus 1$\sigma$ error bars) in bins of 0.15\,dex.
    The black line stands for the \citetalias{Tremonti:2004p1138} fitting,
    while the blue-lines correspond to the \citetalias{Kewley:2008p1394}
    $\pm$0.2\,dex relation.  
    The rectangle encompasses the range in mass and metallicity of the galaxy
    sample of this work.
    \label{fig:sim}
}
\end{figure}

Fig. \ref{fig:sim} shows the result of the simulation, i.e. the
distribution of the mock galaxies in the \mz\ parameter space. We
reproduce the overall shape of the global \mz\ relation assuming a
local \mz\ relation and considering the aperture effect of the SDSS fiber,
over $\sim$5 orders of magnitude in mass and $\sim$1.5 in metallicity. 
The correspondence can be clearly seen by comparing the distribution of points
with the overlaid lines corresponding to the
\citetalias{Tremonti:2004p1138} fit (black) and the \citetalias{Kewley:2008p1394}
$\pm$0.2\,dex relation (blue), for which the agreement is extremely good over
a wide range of masses. The deviations of the \citetalias{Tremonti:2004p1138}
relation at the high-mass end are somewhat expected
given that the choice of this metallicity calibrator has a significant effect on
the $y$-intercept of the \mz\ relation \citepalias{Kewley:2008p1394}.
The result is remarkable considering that we are able to reproduce the global \mz\
relation over a huge dynamical range, using a local \mz\ relation derived from
a galaxy sample with a restricted range in mass (9.2\,$<\log M_{\rm Lum}<$\,11.2)
and metallicity (8.3\,$<$\,12\,+log(O/H)\,$<$\,8.9), indicated by the rectangle
shown in Fig. \ref{fig:sim}.

\section{On the origin of the local and global \mz\ relations}
\label{sec:mzr}

We interpret the local \mze\ relation as the consequence of a more subtle
relation between the mass, metallicity and star formation history with galaxy radius. 
Galaxies are known to have radial gradients in their physical properties
\citep[e.g.][]{Searle:1971p1962,McCall:1985p1243,VilaCostas:1992p322,Zaritsky:1994p333,Bell:2000p4174}.
The fact that gas metallicity increases with mass surface density is just
reflecting the existence of radial metallicity gradients and radial surface
mass density gradients in spirals. 
On the other hand, the \ewha\ is a parameter that scales with the SFR
per unit mass, i.e. the specific SFR (SSFR) and is a proxy of the stellar
birthrate parameter $b$-Scalo, which is the ratio
of the present to past-average SFR \citep{Kennicutt:1994p4173,Kennicutt:1998p3370}.
Multiphase chemical evolution models for spiral discs show that both the
distributions of neutral and molecular gas show maxima that move from the
center outward through the disc as the galaxy evolves. Accordingly, the
maximum of the star formation activity, very high in the central regions at
early times, also moves throughout the disc as the gas in the central region
is very efficiently consumed. This leads to high present day abundances and
low current SSFRs in the center and radial abundance gradients that flatten
with time. The location on the disc of the maxima in the gas distributions at
the present time, which reflects the degree of evolution, is related to the
galaxy morphological type with early type galaxies showing the peak of the
distributions outer than late type galaxies, which implies a higher degree of
evolution and a flatter galaxy for early type galaxies \citep{Molla:1997p4182,Molla:2005p3850}.

The observed trend agrees qualitatively with this scenario: the inverse gradient of \ewha\
vs. $\Sigma_{\rm Lum}$ or metallicity would reflect an evolutionary sequence
across the disc of the galaxies, i.e. lower (inner) values of \ewha\
corresponding to lower SSFR, and vice versa. 
In this scenario, the inner regions of the galaxy form first and faster, increasing the
gas metallicity of the surrounding interstellar medium. As the galaxy evolves
and grows with time, the star-formation progresses radially, creating a radial
metallicity gradients in the disc of spirals. Mass is progressively accumulated
at the inner regions of the galaxy, rising the surface mass density and
creating a bulge, with corresponding high metallicity values but low SSFR (low
\ewha), i.e. an ``inside-out'' galaxy disc growth. 
In such a case, the local \mz\ relation would reflect a more fundamental
relation between mass, metallicity and star-formation efficiency as a function
of radius, equivalent to a local {\em downsizing} effect.

Following this reasoning, the origin of the global \mz\ relation can be
explained as the combined effect of the existence of the local \mz\ relation,
and (as a second order effect) an aperture bias due to the different covering
factors of the SDSS fiber, as suggested by the exercise performed in Sec.\,\ref{sec:global}.
On the other hand, although we cannot reproduce the FMR in our simulation
given that there is no empirical way of determining a SFR of the mock galaxies,
the existence of the FMR could also be 
interpreted as a scaled-up version of the local \mz-SSFR relation.
Furthermore, under the proposed scenario, the flattening of the \mz\ relation
at the high-mass end could be explained as the combined observational effect
of: 1) an intrinsic saturation of the enriched gas due to the low SSFR in the
inner regions of the galaxies; 2) a deviation from linearity of the abundance
gradients in the center of spiral galaxies, i.e. a flattening (or drop) in the
metallicity gradient at the innermost radii as suggested by recent works 
\citep[e.g.][]{RosalesOrtega:2011p4146,Bresolin:2012p4176,Sanchez:2012b} and predicted from
chemical evolution models \citep{Molla:1997p4182}; and 3) the depletion of
the oxygen emission lines in the optical due to high-efficiency of cooling in
high-metallicity (12\,+\,log(O/H)\,$>8.7$), low-temperature ($T_{\rm e}<10^4$\,K)
\hh regions.

\section{Conclusions}
\label{sec:end}

By using IFS of a sample of nearby galaxies we demonstrate the existence of a
{\em local} relation between the surface mass density, gas-phase oxygen
abundance and \ewha\ in $\sim$2000 spatially-resolved \hh regions of the
local Universe. The projection of this distribution in the metallicity
vs. $\Sigma_{\rm Lum}$ plane --the {\em local} \mz\ relation-- shows a tight
correlation expanding over a wide range in this parameter space. A similar
behaviour is seen for the \ewha\ vs. $\Sigma_{\rm Lum}$ relation. In both
projections, the value of \ewha\ is inversely proportional to mass and
metallicity. Notably, the {\em local} \mz\ relation has the same shape as the
global \mz\ relation for galaxies observed by \citetalias{Tremonti:2004p1138}.
We explain the new relation as the combination of: i) the well-known
relationships between both the mass and metallicity with respect to the
differential distributions of these parameters found in typical disc galaxies,
i.e. the {\em inside-out} growth; and ii) the fact that more massive regions
form stars faster (i.e. at higher SFRs), thus earlier in
cosmological times, which can be considered a local {\em downsizing} effect,
similar to the one observed in individual galaxies \cite[e.g.][]{PerezGonzalez:2008p4180}.

We use the local \mz\ relation to reproduce the
global \mz\ relation by means of a simple simulation which
considers the aperture effects of the SDSS fiber at different redshifts. 
We conclude that, the \mz\ relation in galaxies is a scale-up integrated
effect of a local \mz\ relation in the distribution of star-forming regions across
the discs of galaxies, i.e. the relationship is not primary, but obtained from
the sum of a number of local linear relations (and their deviations) with
respect to the galaxy radius. 
Under these premises, the existence of the FMR in galaxies might be explained by
the presence of an intrinsic local \mz-SSFR relation, which relates the
distribution of mass, metallicity and SFR across the galaxy discs, driven
mainly by the history of star formation, within an {\em inside-out} growth
scenario.

\acknowledgments
{\small
\noindent
F.F.R.O. acknowledges the Mexican National Council for Science and Technology
(CONACYT) for financial support under the programme Estancias Posdoctorales y
Sabáticas al Extranjero para la Consolidación de Grupos de Investigación,
2010-2011. 
A.\,I.\,D. thanks the Spanish Plan Nacional de Astronomía programme
AYA2010-21887 C04-03.

}
\vfill



\bibliographystyle{apj}
{\footnotesize
{\scriptsize

}
\end{document}